# Pressure-induced decomposition of $Bi_{14}WO_{24}$

E. Karaca[1], D. Santamaria-Perez[2], A. Otero-de-la-Roza[3], R. Oliva[4], K.S. Rao[5], S.N. Achary[5], C. Popescu[6], D. Errandonea[2,*]

[1]Sakarya University, Faculty of Sciences, Department of Physics, 54050, Sakarya, Turkey

[2]Departamento de Física Aplicada-ICMUV, MALTA-Consolider Team, Universidad de Valencia, Dr. Moliner 50, Burjassot, 46100 Valencia, Spain

[3]Departamento de Química Física y Analítica, Facultad de Química, Universidad de Oviedo, MALTA Consolider Team, 33006, Oviedo, Spain

[4]4Geosciences Barcelona (GEO3BCN), CSIC, Lluís Solé i Sabarís s/n, 08028 Barcelona, Catalonia, Spain

[5]Chemistry Division, Bhabha Atomic Research Centre, Trombay, Mumbai, 400085, India

[6]CELLS-ALBA Synchrotron Light Facility, Cerdanyola del Vallès, 08290 Barcelona, Spain

*Corresponding author: E-mail: daniel.errandonea@uv.es

**Abstract:**

We present a study of the high-pressure behaviour $Bi_{14}WO_{24}$, a high oxide ion conductor member of the $Bi_2O_3$-$WO_3$ binary system. The tetragonal polymorph of $Bi_{14}WO_{24}$ was studied under high-pressure conditions using synchrotron powder X-ray diffraction. It was found that in contrast to isostructural $Bi_{14}CrO_{24}$ and $Bi_{14}MoO_{24}$ which experience a phase transition around 5 GPa, in our study $Bi_{14}WO_{24}$ undergoes an irreversible chemical decomposition into $Bi_2O_3$ and $WO_3$ at 2.85(5) GPa. The pressure dependence of the unit-cell parameters of $Bi_{14}WO_{24}$ was also determined, and hence the linear compressibility along different axes and room-temperature pressure-volume equation of state were derived. Bulk modulus of tetragonal $Bi_{14}WO_{24}$ was found to be 49.8(2.6) GPa, and the linear compressibility of the two crystallographic axes, $\kappa_a$ and $\kappa_c$ were 6.94(2) $10^{-3}$ $GPa^{-1}$ and = 3.73(1) $10^{-3}$ $GPa^{-1}$, respectively. The pressure induced decomposition can be attributed to the favourable increasing density of the system to accommodate the pressure induced stress.

## 1. Introduction

Solid solutions derived from the $Bi_2O_3$-$WO_3$ system exhibit significant versatility. The tungsten-rich phases, like $Bi_2WO_6$ and $Bi_2W_2O_9$, are utilized in the production of technologically relevant photocatalysts, while the bismuth-rich phases, like $Bi_{14}WO_{24}$, serve as solid electrolytes that conduct $O^{2-}$ ions, demonstrating exceptional stability under low $O_2$ partial pressures and high ionic conductivities [1]. In compositions that are rich in bismuth, the structures exhibit a strong correlation with δ-$Bi_2O_3$, a fluorite prototype with 25% oxygen vacancies and one of the most promising solid electrolyte materials. The investigation of the $Bi_2O_3$-$WO_3$ system has been initiated in 1938 with the identification of the mineral russellite, a bismuth tungstate with the chemical formula $Bi_2WO_6$ [2]. W. Zhou has reported a comprehensive review on structure and properties of different phases identified in the $Bi_2O_3$-$WO_3$ system [3]. Among these phases, $Bi_{14}WO_{24}$ is one of the ordered phases in the $Bi_2O_3$ rich region exhibit temperature induced phase transition [4]. At ambient condition, $Bi_{14}WO_{24}$ crystallizes in the tetragonal *I*4/*m* space group and it also exists as a second polymorph described by the monoclinic space group *C*2/*m* [4]. $Bi_{14}WO_{24}$ is isostructural to $Bi_{14}MoO_{24}$ and $Bi_{14}CrO_{24}$, and all of them undergo phase transition from the tetragonal structure to the monoclinic structure at low temperature. The transition temperatures for $Bi_{14}WO_{24}$, $Bi_{14}MoO_{24}$ and $Bi_{14}CrO_{24}$ are as 306 K, 295 K, and 200 K, respectively. $Bi_{14}MoO_{24}$ and $Bi_{14}CrO_{24}$ have been studied also under high-pressure (HP) conditions [5,6,7]. The two compounds undergo a tetragonal-tetragonal transition around 5 GPa and subsequently the structure becomes monoclinic at a pressure that depends on the non-hydrostatic stresses present in the experiments [7]. All the structural transitions are related to the increasing density and efficient packing of the ions in the structure. Thus, it is expected that $Bi_{14}WO_{24}$ may also undergoes similar transition or some other denser phases with pressure as it exhibits temperature induced transition just near the ambient temperature. However, to the best of our knowledge no study on high-pressure behaviour of $Bi_{14}WO_{24}$ is available in literatures till date. In the aim to study the stability and possible phase transition, and high-pressure studies on the $Bi_{14}WO_{24}$ were undertaken experimentally and theoretically.

Herein we will report the first high-pressure study on the tetragonal phase of tetradecabismuth tungsten tetracosaoxide, $Bi_{14}WO_{24}$, which is described by space group *I*4/*m*. The crystal structure is represented in Figure 1. The tungsten atoms can be interpreted as tetrahedrally coordinated with O atoms, exhibiting significant orientational disorder. Hence this also be described as octahedrally coordinated W with two unoccupied oxygens sites, one axial and one equatorial oxygen site unoccupied. Thus, this six-coordination polyhedron can be represented as $WO_4[]_2$. The $WO_4[]_2$ octahedra are isolated from each other and positioned at the corners and centre of the unit cell. Additionally, it may be noted here that none of the oxygen atoms of $WO_4[]_2$ units are linked to $BiO_4$ polyhedra. Another peculiarity of the tetragonal structure of $Bi_{14}WO_{24}$, is related to the role played by the lone pair of electrons (LEP) of bismuth. Indeed, the crystal structure that can be characterized as a Bi–O framework, consisting of interconnected [Bi(LEP)$O_4$] square pyramids featuring an apical LEP, as well as trigonal bipyramids with an equatorial LEP. The trigonal pyramids bear resemblance to the coordination polyhedron of bismuth atoms found in α-$Bi_2O_3$. Additionally, the square pyramids are aggregated to form clusters with the composition $Bi_6O_8$, where the eight oxygen (O) atoms create a distorted cubic arrangement (see Fig. 1). The loosely arranged lattice provides sufficient space for larger $WO_4[]_2$ octahedra that are orientationally disordered within the structure. This fact and the presence of stereochemically active $6s^2$ LEP associated with Bi atoms, suggest that the crystal structure could be easily modified under a compression of just a few GPa [8].

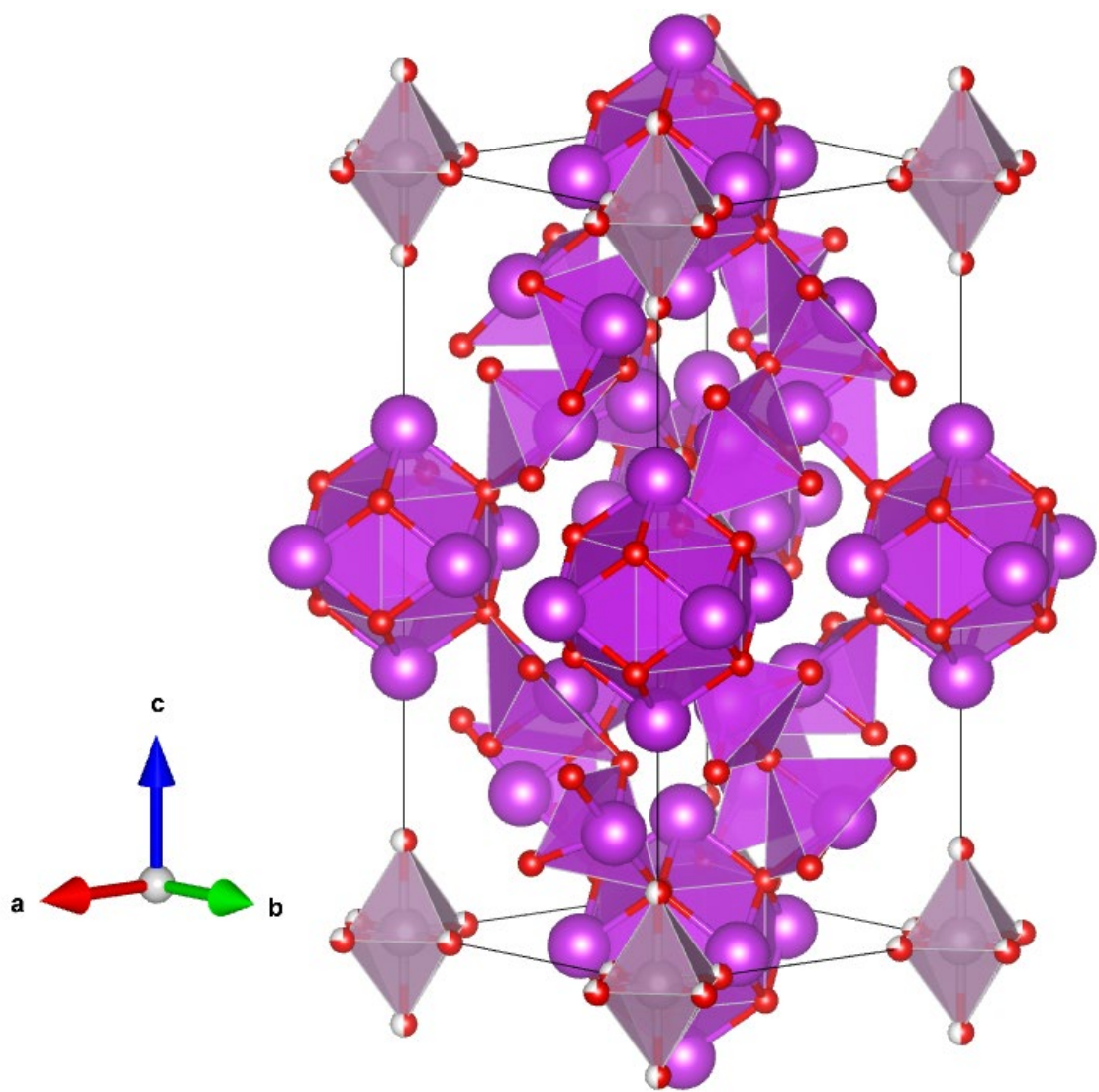

**Figure 1:** Schematic representation of the tetragonal structure of $Bi_{14}WO_{24}$ described by space group $I4/m$. We used grey colour for W, purple for Bi, and red for O atoms (partial coloured indicating partial occupancy). The polyhedral units described in the text are shown.

In this work, we investigated the HP behaviour of the tetragonal polymorph of $Bi_{14}WO_{24}$ using a diamond-anvil cell (DAC) and synchrotron-based powder X-ray diffraction (XRD). We found that, in contrast with isomorphic $Bi_{14}MoO_{14}$ or $Bi_{14}CrO_{24}$ compounds, $Bi_{14}WO_{24}$ chemically decomposes at a pressure of 2.85 GPa. The decomposition is also found to be irreversible. This finding is supported by density-functional theory calculations. Additionally, we obtained the pressure dependence of unit-cell parameters and determined a room-temperature (RT) isothermal pressure-volume (P-V) equation of state (EoS) and bulk modulus. We also performed a comparative analysis by comparing the bulk moduli of other of members of the $Bi_2O_3$-

WO$_3$ binary system. It revealed that the increase in the the content of WO$_3$ increase of the bulk modulus of compounds in Bi$_2$O$_3$-WO$_3$ system.

## 2. Methods

### 2.1 Experiments

Powder samples of Bi$_{14}$WO$_{24}$ were obtained through a solid-state reaction involving stoichiometric mixtures of Bi$_2$O$_3$ and WO$_3$. Both oxides were obtained from Alfa Aesar with 99% purity. Before the synthesis, Bi$_2$O$_3$ underwent heating at 973 K for 6 hours in air, followed by a cooling process to RT at a rate of 2 °C/min. WO$_3$ was subjected to heating at 573 K overnight. The heating of Bi$_2$O$_3$ was carried out to was carried out to remove any hydroxide or carbonate impurities in the Bi$_2$O$_3$ [9] and the heating of WO$_3$ was performed to remove any adsorbed moisture [10]. The preheated reactants were then combined in stoichiometric proportions (7 Bi$_2$O$_3$ + WO$_3$) and thoroughly mixed using an agate mortar and pestle. The resulting mixture was pressed into pellets measuring 10 mm in diameter and 2 mm in thickness. These pellets were subsequently placed in a platinum boat and heated at 973 K for 12 hours, the rate of the heating and cooling processes being 2 °C/min. The homogeneous pale-yellow product obtained post-heat treatment was characterized using powder XRD. It was identified as the tetragonal polymorph (space group *I*4/*m*) previously reported by Crumpton *et al.* [4].

Angle-dispersive powder XRD under high pressure was conducted at the BL04-MSPD beamline of the ALBA-CELLS synchrotron [11]. These experiments utilized a DAC, with diamond culets of 500 μm in diameter. For the gasket material, a stainless-steel disk measuring 200 μm in thickness was pre-indented to a final thickness of 40 μm, featuring a centrally drilled hole with a diameter of 200 μm. The pressure medium employed was a 4:1 mixture of methanol and ethanol. Pressure measurements were done with an accuracy of ±0.05 GPa, based on the equation of state for copper [12]. The selected pressure medium ensures quasi-hydrostatic conditions up to at least 10.5 GPa [13]. XRD experiments were carried out using a monochromatic X-ray beam with a wavelength of 0.4246 Å, focused to a spot size of 30 μm × 30 μm using Kirkpatrick–Baez mirrors. Two-dimensional (2D) XRD patterns were recorded using a Rayonix SX165 charge-coupled device, calibrated with lanthanum hexaboride as a standard. The detector-sample distance was 300.15 mm. The resulting 2D diffraction images were processed into intensity versus 2θ XRD patterns utilizing DIOPTAS [14], followed by Rietveld refinements or Le Bail fits conducted using the FullProf suite [15].

*2.2 Computer simulations*

DFT calculations were performed utilizing Quantum ESPRESSO [16], applying the generalised gradient approximation with the Perdew-Burke-Ernzerhof parametrisation [17] and employing using ultrasoft pseudopotentials [18], which included 15 valence electrons for bismuth (Bi), 14 for tungsten (W), and 6 for oxygen (O). High cut-off values were chosen for the plane-wave expansion of the Kohn-Sham states (120 Ry) and for the electron density (1200 Ry). A uniform k-point grid of 8 × 8 × 8 was employed for all structures, ensuring a convergence of approximately 0.1 mRy in the total energy. From calculations, we determined the pressure dependence of the volume of $Bi_{14}WO_{24}$ and other $Bi_2O_3$ – $WO_3$ binary compounds and the enthalpy as function of pressure for $Bi_{14}WO_{24}$ and the decomposition products $Bi_2O_3$ and $WO_3$. The pressure (P) was obtained for each compound after the optimization of the structures at selected volumes from energy versus volume (V) plots as the derivative of the total energy with respect to volume. The enthalpy (H) as function of pressure was then obtained for each structure as H = E + P × V.

## 3. Results and discussion

Figure 2 presents a series of powder XRD patterns recorded at various pressures during the compression process up to 2.85 GPa and subsequent decompression to ambient pressure. It is observed that all XRD patterns obtained at pressures below 2.85 GPa correspond to the known tetragonal phase of $Bi_{14}WO_{24}$ described in the introduction. This conclusion is supported by the Rietveld refinements shown in Fig. 2. The refinements and the R-values (given in the figure) support the agreement of the structural model with the experimental data. Notably, at 2.85 GPa, significant alterations in the XRD pattern are evident (see Fig. 2), which signify the occurrence of changes in the sample. The changes in the XRD pattern are non-reversible as shown in Fig. 2. To try to explain the changes in XRD patterns we considered the possibility of the occurrence of a structural phase transition. We found that the XRD pattern measured at 2.85 GPa cannot be assigned to the known monoclinic structure of $Bi_{14}WO_{24}$, described by space group $C2/m$ [4]. We also found that the XRD pattern collected at 2.85 GPa cannot be explained by any of the HP phases found in $Bi_{14}CrO_{24}$ or in $Bi_{14}MoO_{24}$ [5,6,7]. In our efforts to identify a candidate structure for the possible HP phase, we employed the DICVOL routine to index the Bragg reflections obtained from X-ray diffraction measurements conducted at 2.85 GPa. However, the data could not be indexed to a single phase, even when considering a triclinic symmetry, which is the least symmetric among the 14 three-dimensional Bravais lattices.

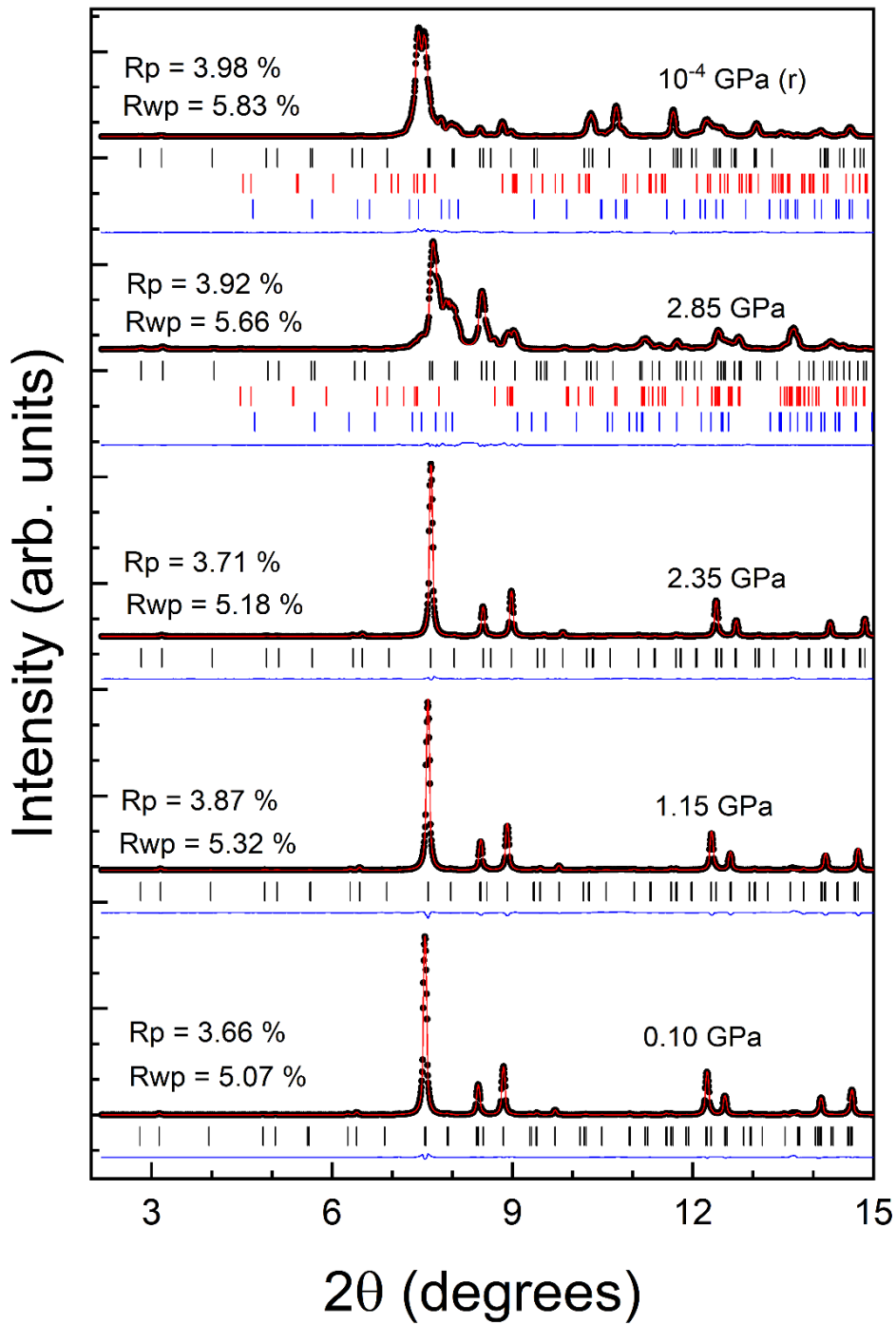

**Figure 2:** Selection of XRD patterns measured at different pressures (indicated in the figure). Black symbols are used for the measured XRD patterns. Blue (green) lines are used for the calculated patterns (residuals). For 0.10, 1.15, and 2.35 GPa Rietveld refinements were performed. For 2.85 and $10^{-4}$ GPa Le Bail fits were performed. The XRD pattern obtained after pressure release is denoted by (r, from "recovered"). R-values are given in the figure. Black ticks correspond to the Bragg peaks of $Bi_{14}WO_{24}$, and red (blue) ticks correspond to the Bragg peaks of $Bi_2O_3$ ($WO_3$).

Pressure-induced decomposition is not an unusual phenomenon. Subsequently we consider the possibility of a pressure-induced decomposition of $Bi_{14}WO_{24}$ at 2.85 GPa. We found that the XRD pattern measured at this pressure could be explained as the coexistence of $Bi_{14}WO_{24}$ with $\alpha$-$Bi_2O_3$ [19] and the monoclinic structure of $WO_3$ stable from 1 to 20 GPa [20]. This conclusion is sustained by the Le Bail fit presented in Fig. 2, which assigns all the peaks measured at 2.85 GPa either to tetragonal $Bi_{14}WO_{24}$, $\alpha$-$Bi_2O_3$, or $WO_3$. In this case, a Rietveld refinement was not performed because of the presence of preferred orientations. The fitting of all peaks by the Le Bail methods indicates that the XRD experiments are consistent with the onset of a decomposition of $Bi_{14}WO_{24}$ into its constituent oxides between 2.35 GPa (the last pressure where we observed only $Bi_{14}WO_{24}$) and 2.85 GPa (the first pressure where we observed decomposition). This phenomenon has been observed in other oxides under compression; for instance, in $Cu_3V_2O_8$ [21] and $La_2Hf_2O_7$ [22]. The occurrence of chemical decomposition is consistent with the fact that the process is irreversible as shown by the XRD pattern measured at room conditions after decompression (see Fig. 2). The decomposition of $Bi_{14}WO_{24}$ into 7 $Bi_2O_3$ + $WO_3$ is also supported by enthalpy vs. pressure calculations which we performed to test the stability of $Bi_{14}WO_{24}$ at different pressures. Figure 3 shows that according to our calculations the enthalpy of the constituent oxides becomes smaller than that of $Bi_{14}WO_{24}$ beyond 2.4 GPa, indicating that the decomposition of this compound is thermodynamically favoured [23,24] at the pressure where it is found experimentally.

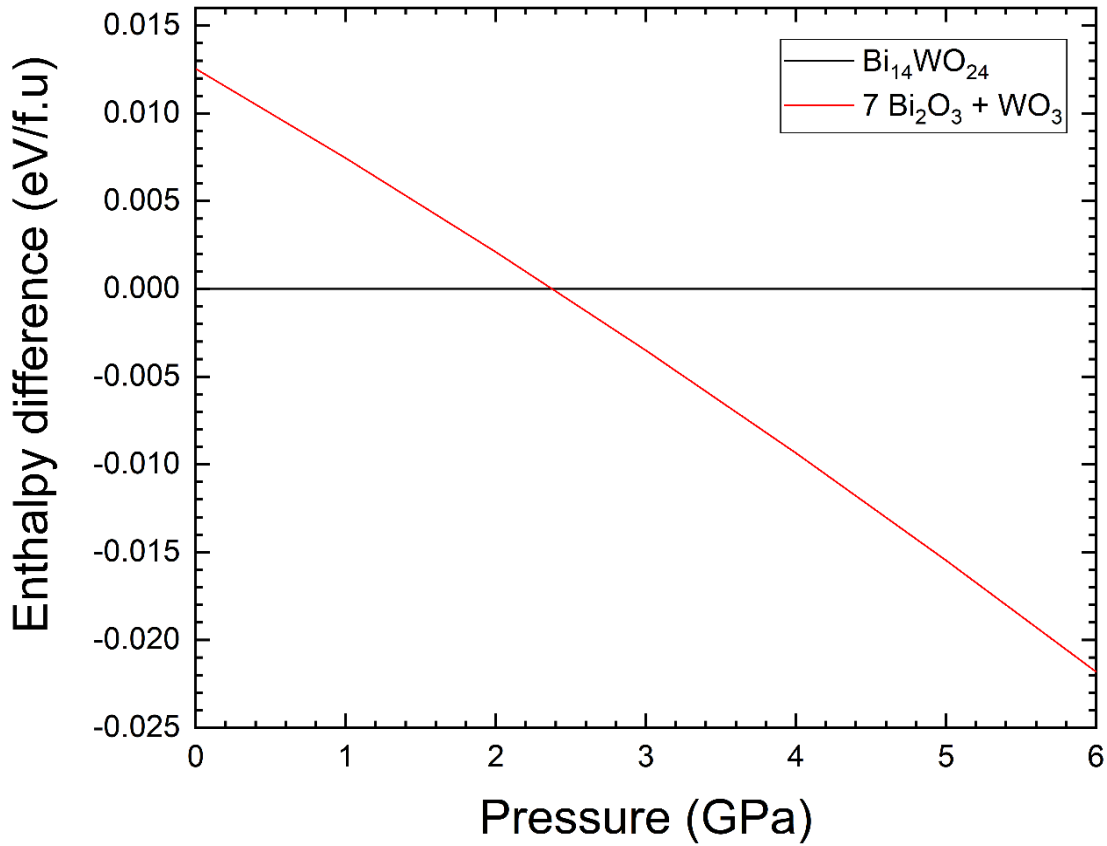

**Figure 3:** Enthalpy difference between the decomposition products (7 $Bi_2O_3$ + $WO_3$) and tetragonal $Bi_{14}WO_{24}$.

The dissociation of a compound into its constituent components under high pressure is predominantly influenced by atomic diffusion, which tends to be significant only at elevated temperatures. It is possible that, even though the enthalpy of the resulting products may be lower than that of the parent compound at high pressures, decomposition may not occur. This is because the volume per formula unit of the products can exceed that of the parent compound. The presence of a potential energy barrier along the pathway to the product phases can result in kinetic hindrance, ultimately blocking the decomposition [22]. However, in the case studied here, the volume factors are conducive to decomposition. Using the volumes per formula unit for $Bi_{14}WO_{24}$ ($V_{BWO}$) [4], $Bi_2O_3$ ($V_{BO}$) [19], and $WO_3$ ($V_{WO}$) [20], as determined from our XRD analysis at 2.85 GPa, we obtained that the fractional reduction in volume upon decomposition, calculated as ($V_{BWO}$ - ($V_{WO}$ + 7 $V_{BO}$)) / $V_{BWO}$, is approximately 4.9%. Thus, higher density as well as lower enthalpy of the products at high pressure favour the decomposition in $Bi_{14}WO_{24}$. It may be noted that $W^{6+}$ has more preference for octahedral coordination compared to $Cr^{6+}$ or $Mo^{6+}$, and that is favoured while compression of the highly orientationally disordered $WO_4[]_2$. In $WO_3$ $W^{6+}$ is in octahedral coordination. Thus, the compression might lead to phase separation $\alpha$-$Bi_2O_3$ and $WO_3$ phases to accommodate the stress in the $Bi_6O_8$ and $WO_4[]_2$ units. We would like to note here that there might be more explanations to the observed phenomenon, such as different ionic between Cr, Mo, and W. However, a more detailed discussion why decomposition occurs in $Bi_{14}WO_{24}$ while phase transitions occur in $Bi_{14}CrO_{24}$ and $Bi_{14}MoO_{24}$ is beyond the scope of the present work.

From the XRD patterns we obtained the pressure dependence of the unit cell parameters of $Bi_{14}WO_{14}$. The results are given in Table 1 and represented in Figure 4. In Figure 5 we represent the pressure dependence of the unit-cell volume. The lattice parameters follow a linear dependence with pressure. The linear compressibility of each of the axes is $\kappa_a$ = 6.94(2) $10^{-3}$ $GPa^{-1}$ and $\kappa_c$ = 3.73(1) $10^{-3}$ $GPa^{-1}$. This means that compressibility is anisotropic, being the a-axis the most compressible one. This fact can be clearly seen in the inset of Fig. 4, where the axial ratio c/a is represented. It increases from a value of 1.99 at 0 GPa to 2.01 at 2.55 GPa. The results from experiments are supported by present density-functional theory calculations. The results from DFT calculations are included for comparison in Fig. 4. Calculations underestimate the c parameter by 1.2% and the a parameter by 0.5%, which could be considered as a good agreement. The pressure dependence obtained from calculations is similar to that obtained from experiments.

| Pressure (GPa) | a (Å) | c (Å) | V (Å$^3$) |
|---|---|---|---|
| $10^{-4}$ | 8.730(3) | 17.361(6) | 1323.0(1.3) |
| 0.10(5) | 8.716(3) | 17.354(6) | 1318.4(1.3) |
| 0.15(5) | 8.715(3) | 17.348(6) | 1317.5(1.3) |
| 0.60(5) | 8.688(3) | 17.320(6) | 1307.4(1.3) |
| 1.00(5) | 8.661(3) | 17.293(6) | 1297.2(1.3) |
| 1.15(5) | 8.651(3) | 17.282(6) | 1293.4(1.3) |
| 1.35(5) | 8.638(3) | 17.270(6) | 1288.7(1.3) |
| 1.80(5) | 8.612(3) | 17.237(6) | 1278.4(1.3) |
| 2.15(5) | 8.599(3) | 17.219(6) | 1273.3(1.3) |
| 2.35(5) | 8.582(3) | 17.206(6) | 1267.2(1.3) |
| 2.55(5) | 8.571(3) | 17.199(6) | 1263.4(1.3) |

**Table 1:** Unit-cell parameters of $Bi_{14}WO_{24}$ at room temperature as a function of pressure.

The pressure dependence of the volume (shown in Fig. 5) was fitted with a third-order Birch-Murnaghan EoS [25]. From the experiments we obtained the unit-cell volume at zero pressure $V_0$ = 1322.0(6) Å$^3$, the bulk modulus at zero pressure $K_0$ = 49.8(2.6) GPa, and its pressure derivative $K_0'$ = 5.5(2.3). The value for the pressure derivative agrees within one standard deviation with the implied value of a second-order truncation of the EoS ($K_0'$ = 4). Assuming this order for the fit, we obtained $V_0$ = 1321.7(8) Å$^3$, the bulk modulus at zero pressure $K_0$ = 51.5(7) GPa. Both fits give a very similar pressure dependence of the volume. Indeed, the fits are nearly indistinguishable in Figure 5. The inset of the figure shows a plot of $K_0'$ versus $K_0$. The result of the second-order fit is within the 68.3 % confidence ellipse of the third order fit, indicating that the fits are equivalent. The maximum difference in pressure between the fits and the experiments is 0.07 GPa. From DFT calculations we obtained $V_0$ = 1297.3(2) Å$^3$, $K_0$ = 52.1(2) GPa, and $K_0'$ = 3.9(4). The calculated volume is 2% smaller than the experimental value. The calculated $K_0$ and $K_0'$ agree with the results obtained from experiments (see the inset of Fig. 5).

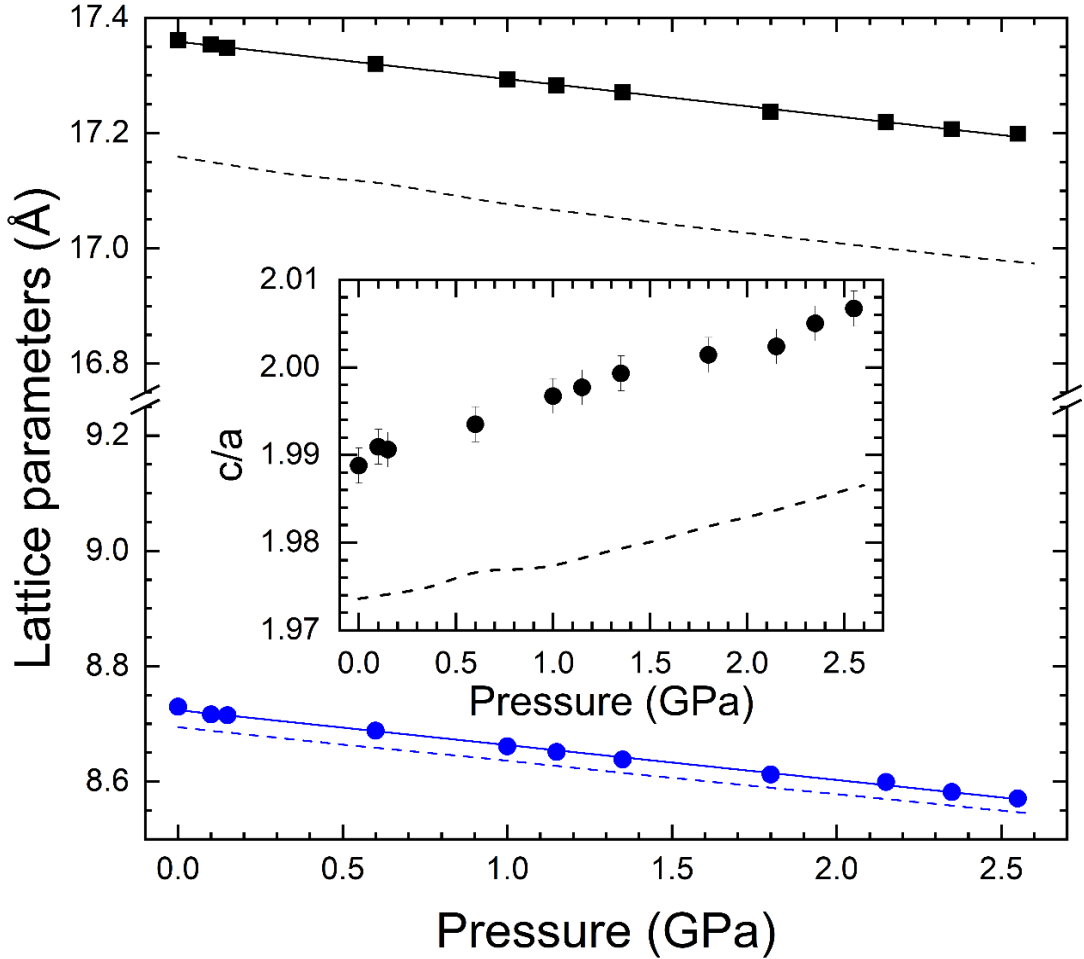

**Figure 4:** Pressure dependence of the unit-cell parameters of $Bi_{14}WO_{24}$. Symbols correspond to experimental results (error bars are smaller than symbols) and lines are linear fits. The inset shows the pressure dependence of the axial ratio c/a.

The value of the bulk modulus of $Bi_{14}WO_{24}$ is slightly smaller than the bulk modulus of $Bi_{14}CrO_{24}$, 56(6) GPa [5], and $Bi_{14}MoO_{24}$, 67(7) GPa [6]. We will next compare it with the bulk modulus of different compounds of the $Bi_2O_3$ – $WO_3$ binary system. The only other compound of this system studied in experiments up to now is $Bi_2WO_6$ and it has a bulk modulus of 79(4) GPa [26]. The increase of the bulk modulus in $Bi_2WO_6$ is probably related to the increase of the amount of $WO_3$ in the compound, 12.5 % in $Bi_{14}WO_{24}$ and 50% in $Bi_2WO_6$. This hypothesis is fully reasonable because in one hand Bi-O bonds are much more compressible [19] than W-O bonds [20] and on the other hand, the increase of the $WO_3$ content in the compound reduces the concentration of Bi LEP, whose presence favours compressibility [8].

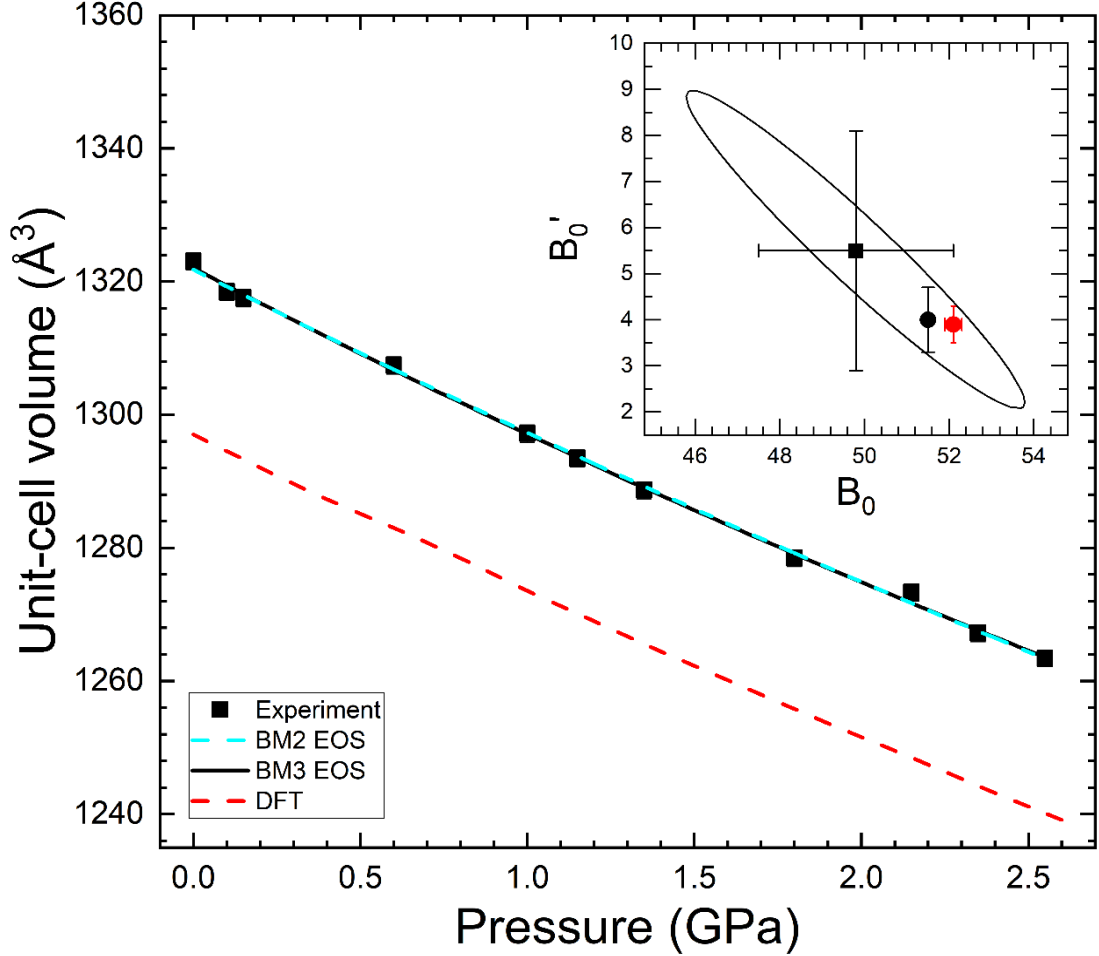

**Figure 5:** Pressure dependence of the unit-cell volume. Symbols are from experiments. The black solid (cyan dashed) line is the third (second) order equation of state obtained from experiments and described in the text. The red dashed line is the third-order equation of state obtained from DFT calculations. The inset represents $B_0'$ versus $B_0$. The black circle (square) is the result of the second (third) order equation of state. The red circle is the result obtained from DFT calculations using a third-order equation of state. The 68.3% confidence ellipse from the third-order fit is shown.

To test our hypothesis, we have calculated the pressure dependence of the volume of $Bi_4WO_9$ (33 % of $WO_3$) and $Bi_2W_2O_9$ (66 % of $WO_3$). The first compound has not been studied yet under compression. The second one was studied by Raman spectroscopy and phase transitions were found at 2.8 and 4.8 GPa [27], but the pressure dependence of the volume remains unknown. From our calculations we determined the EoS for both compounds. The obtained parameters are given in Table 2, compared with present and previous experimental results from $Bi_{14}WO_{24}$ and $Bi_2WO_6$. For the last one, for completeness, we also include results from DFT calculations we performed. Our DFT simulations agree with experiments. Notice that for all compounds the pressure derivative of the bulk modulus is close to 4.0, then bulk moduli can be directly compared.

| Compound | Content of WO$_3$ | EoS | $V_0$ (Å$^3$) | $K_0$ (GPa) | $K'_0$ |
|---|---|---|---|---|---|
| Bi$_{14}$WO$_{24}$ | 12.5% | Exp. 3$^{rd}$ order | 1322.0(6) | 49.8(2.6) | 5.5(2.3) |
|  |  | Exp. 2$^{nd}$ order | 1321.7(8) | 51.5(7) | 4.0 |
|  |  | DFT 3$^{rd}$ order | 1297.3(2) | 52.1(2) | 3.9(4) |
| Bi$_4$WO$_9$ | 33% | DFT 3$^{rd}$ order | 1757.8(3) | 74(1) | 4.2(2) |
| Bi$_2$WO$_6$ | 50% | Exp. 2$^{nd}$ order[a] | 486.5(2) | 79(4) | 4.0 |
|  |  | DFT 3$^{rd}$ order | 485.2(2) | 78(1) | 4.1(2) |
| Bi$_2$W$_2$O$_9$ | 66.7% | DFT 3$^{rd}$ order | 697.6(2) | 131.5(9) | 3.6(3) |

**Table 2:** Zero-pressure volume, bulk modulus and its zero-pressure derivative for different Bi$_2$O$_3$ – WO$_3$ binary compounds. [a]Taken from Lis *et al.* [26].

The agreement between experiments and DFT simulations for Bi$_{14}$WO$_{24}$ and Bi$_2$WO$_6$ make us confident in the results we obtained from DFT simulations we performed for Bi$_4$WO$_9$ and Bi$_2$W$_2$O$_9$. Our results show that the bulk modulus of members of the Bi$_2$O$_3$ – WO$_3$ system increases following the sequence Bi$_{14}$WO$_{24}$ (12.5% WO$_3$) < Bi$_4$WO$_9$ (33% WO$_3$) < Bi$_2$WO$_6$ (50% WO$_3$) < Bi$_2$WO$_9$ (66.7% WO$_3$). This result supports our hypothesis that the bulk modulus is expected to increase as the content of WO$_3$ increases. This is also consistent with the fact that the least dense polymorph of Bi$_2$O$_3$ has a bulk modulus of 33(3) GPa [28] and the densest polymorph of WO$_3$ has a bulk modulus of 200(10) GPa at 0 GPa [20]. In fact, we have noticed that the bulk modulus of all members of the Bi$_2$O$_3$ – WO$_3$ binary family, including end members, as a first approximation follows a linear dependence with the WO$_3$ content. This can be seen in Figure 6 where we represent the bulk moduli versus the WO$_3$ content. The linear fit to the six data points is $K_0$ (GPa) = 29(5) + 150(12) x, where x is the fraction of WO$_3$ in the compound. The obtained relationship can be used to estimate the bulk modulus of other compound of the Bi$_2$O$_3$ – WO$_3$ family, such as Bi$_6$WO$_{12}$ (25% WO$_3$), Bi$_{14}$W$_2$O$_{27}$ (22.2% of WO$_3$), and Bi$_{12}$WO$_{21}$ (14.28% WO$_3$) for which a bulk modulus of 66(15), 62(14), and 60(13) GPa, respectively, is predicted.

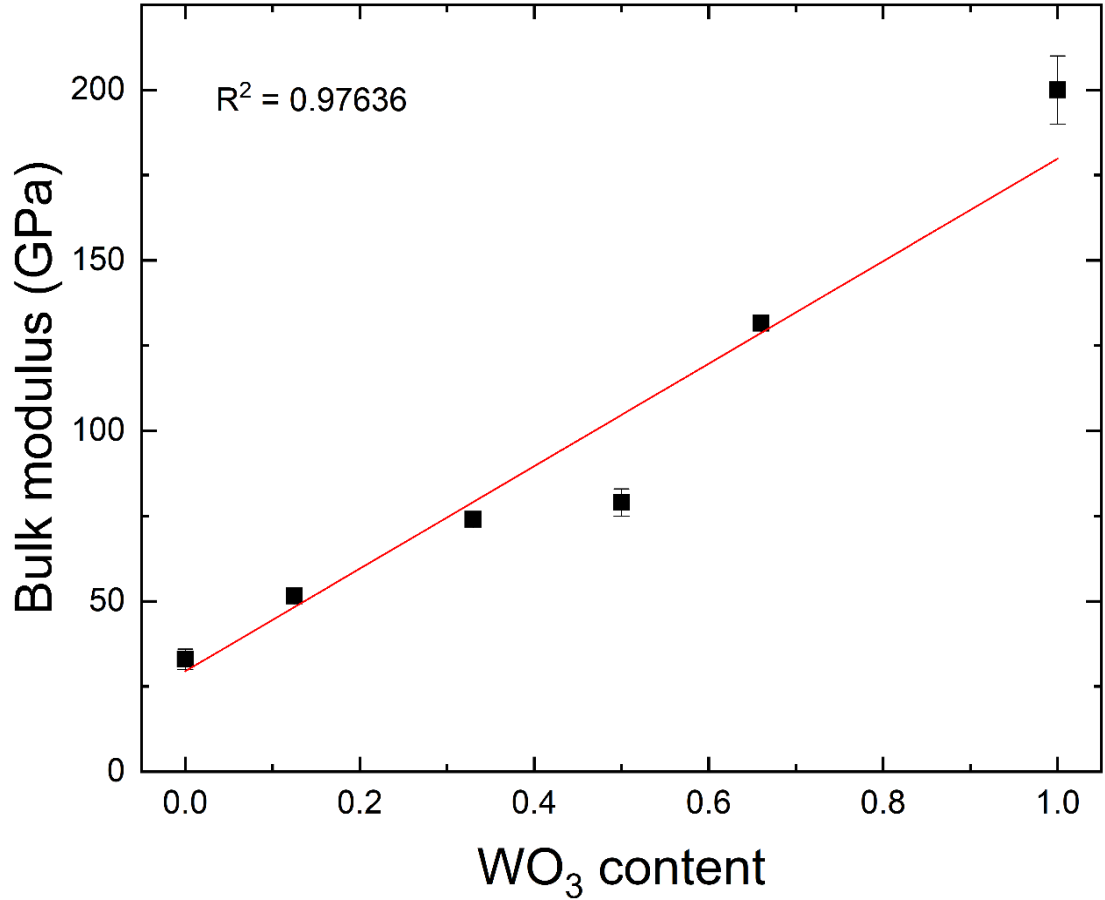

**Figure 6:** Bulk modulus of a compound of the $Bi_2O_3$ -$WO_3$ binary family versus the content of $WO_3$ in %. The data is taken from Table 2. The red line is the linear fit described in the text. The coefficient of determination of the fit $R^2$ is given in the figure.

## 4. Conclusions

In this work we reported a high-pressure synchrotron powder x-ray diffraction study of the tetragonal polymorph of $Bi_{14}WO_{24}$. We conclude that, in our experiment, this compound decomposes into 7 $Bi_2O_3$ + $WO_3$ at 2.85 GPa. The decomposition is irreversible. According to our density-functional theory calculations, the decomposition is favoured by the decrease of the total enthalpy and volume of the decomposition products in comparison to $Bi_{14}WO_{24}$. This conclusion is also supported by the fact that at 2.85 GPa the volume per formula unit of the products (7 $Bi_2O_3$ + $WO_3$) exceeds that of the parent compound ($Bi_{14}WO_{24}$). We have also obtained information of the anisotropic compressibility of the studied compound as well as a room-temperature pressure-volume equation of state. The bulk modulus of $Bi_{14}WO_{24}$ has a value very close to that of the same parameter in isostructural $Bi_{14}CrO_{24}$ and $Bi_{14}MoO_{24}$ with a value of 51.5(7) GPa. The compressibility is mainly driven by compressibility of Bi-O bonds. By comparing $Bi_{14}WO_{24}$ with other members of the $Bi_2O_3$ – $WO_3$ binary family we found that $Bi_{14}WO_{24}$ is one of the most compressible compounds of the family. Moreover, we found that the value of the bulk modulus follows a linear dependence with the content of $WO_3$ in the compound.

**Author Contributions**

D. Errandonea conceived the project. E. Karaca, D. Santamaria-Perez, A. Otero-de-la-Roza, R. Oliva, A.K. Rao, S.N. Achary, C. Popescu, and D. Errandonea carried out the investigation and formal analysis. All authors participated in writing and editing the manuscript. All authors have given approval to the final version of the manuscript.

**Declaration of competing interest**

The authors declare that they have no known competing financial interests or personal relationships that could have appeared to influence the work reported in this paper.

**Data Availability**

The data that support the findings of this study are available from the corresponding author upon reasonable request.

**Acknowledgements**

The authors gratefully acknowledge the financial support thank the financial support from the Spanish Ministerio de Ciencia, Innovación y Universidades (DOI: 10.13039/501100011033) under Projects PID2022-138076NB-C41/44, PID2021-125518NB-I00, CNS2023-144958, and RED2022-134388-T. We would also like to thank the financial support of Generalitat Valenciana through grants PROMETEO CIPROM/2021/075-GREENMAT, CIAICO/2021/241 and MFA/2022/007. This study forms part of the Advanced Materials program and is supported by MCIN with funding from European Union Next Generation EU (PRTR-C17.I1) and by the Generalitat Valenciana. The authors thank ALBA for providing beamtime under experiment No. 2022025712.